\documentclass[12 pt]{article}
\usepackage{xcolor}
\usepackage{graphicx}
\usepackage{epstopdf}
\usepackage{generic}
\usepackage{cite}
\usepackage{amsmath,amssymb,amsfonts}
\usepackage{algorithmic}
\usepackage{array}
\usepackage{authblk}
\usepackage{lineno}  
\newenvironment{keywords}{\par\textbf{Keywords: }}{\par}
\usepackage{textcomp}
\usepackage{siunitx}
\usepackage{subcaption} 
\usepackage{nameref,hyperref}
\usepackage{cleveref}
\usepackage{microtype} 
\def\BibTeX{{\rm B\kern-.05em{\sc i\kern-.025em b}\kern-.08em
    T\kern-.1667em\lower.7ex\hbox{E}\kern-.125emX}}
\begin{document}

\newcommand{\RNum}[1]{\uppercase\expandafter{\romannumeral #1\relax}}
\title{Explainable Parallel CNN-LSTM Model for Differentiating Ventricular Tachycardia from Supraventricular Tachycardia with Aberrancy in 12-Lead ECGs}
%
\author[a]{Zahra Teimouri-Jervekani}
\author[b]{Fahimeh Nasimi \thanks{Corresponding author: f.nasimi@khc.ui.ac.ir}}
\author[c]{Mohammadreza Yazdchi}
\author[c]{Ghazal MogharehZadeh}
\author[c]{Javad Tezerji}
\author[d]{Farzan Niknejad Mazandarani}
\author[e]{Maryam Mohebbi}
\affil[a]{Cardiac Rehabilitation Research Center, Cardiovascular Research Institute, Department of Cardiology, School of Medicine, Isfahan University of Medical Sciences, Isfahan, Iran.}
\affil[b]{Department of Computer Science, Khansar Campus, University of Isfahan, Isfahan, Iran.}
\affil[c]{Department of Biomedical Engineering, Faculty of Engineering, University of Isfahan, Isfahan, Iran.}
\affil[d]{Department of Electrical, Computer and Biomedical Engineering, Toronto Metropolitan University, Toronto, Canada.}
\affil[e]{Department of Biomedical Engineering, Faculty of Electrical Engineering, K.N. Toosi University of Technology, Tehran, Iran.}

\maketitle

\newpage
\begin{abstract}
Background and Objective: Differentiating wide complex tachycardia (WCT) is clinically critical yet challenging due to morphological similarities in electrocardiogram (ECG) signals between life-threatening ventricular tachycardia (VT) and supraventricular tachycardia with aberrancy (SVT-A). Misdiagnosis carries fatal risks. We propose a computationally efficient deep learning solution to improve diagnostic accuracy and provide model interpretability for clinical deployment.  

Methods: A novel lightweight parallel deep architecture is introduced. Each pipeline processes individual ECG leads using two 1D-CNN blocks to extract local features. Feature maps are concatenated across leads, followed by LSTM layers to capture temporal dependencies. Final classification employs fully connected layers. Explainability is achieved via Shapley Additive Explanations (SHAP) for local/global interpretation. The model was evaluated on a 35-subject ECG database using standard performance metrics.  

Results: The model achieved $95.63\%$ accuracy ($95\%$ CI: $93.07–98.19\%$), with sensitivity=$95.10\%$, specificity=$96.06\%$, and F1-score=$95.12\%$. It outperformed state-of-the-art methods in both accuracy and computational efficiency, requiring minimal CNN blocks per pipeline. SHAP analysis demonstrated clinically interpretable feature contributions.  

Conclusions: Our end-to-end framework delivers high-precision WCT classification with minimal computational overhead. The integration of SHAP enhances clinical trust by elucidating decision logic, supporting rapid, informed diagnosis. This approach shows significant promise for real-world ECG analysis tools.

\end{abstract}
%
%



\section{Introduction}
\label{sec1}
\begin{figure}[htbp]
        \raggedright 

  \begin{subfigure}[t]{\textwidth} 
        \raggedright 

    \includegraphics[width=\textwidth]{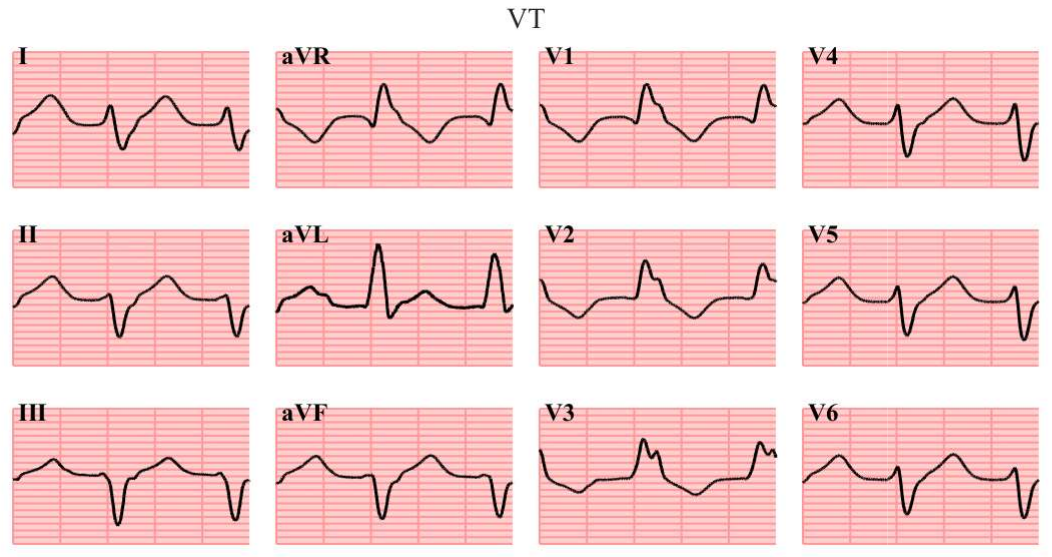}

    \label{fig:top}
  \end{subfigure}
  
  \vspace{0.5cm} 
  
  \begin{subfigure}[t]{\textwidth} 
        \raggedright 

    \includegraphics[width=\textwidth]{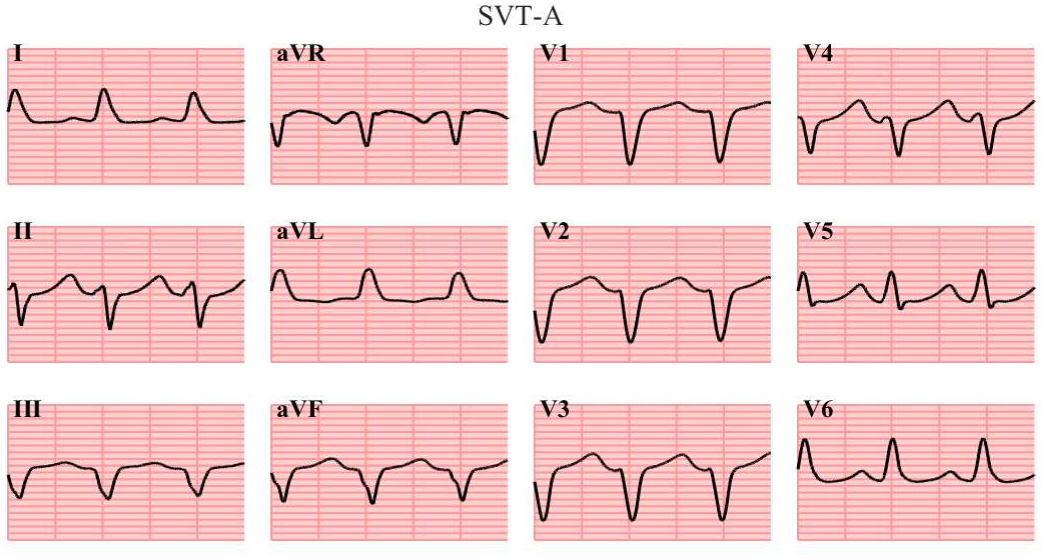}

    \label{fig:bottom}
  \end{subfigure}
  
  \caption{Top) 12-Lead ECG signal with VT. Bottom) 12-Lead ECG signal with SVT-A} 
  \label{fig:both}
\end{figure}

The electrocardiogram (ECG) is a crucial biosignal for cardiologists, providing essential information about the heart's electrical activity. Early detection and discrimination of cardiac arrhythmias can significantly improve patient outcomes and quality of life. One of the main uses of ECG is the differentiation of wide complex tachycardia(WCT).\\
Three primary mechanisms that can lead to WCT are supraventricular tachycardia with intraventricular aberrant conduction, supraventricular tachycardia that conducts to the ventricles via an accessory pathway, and ventricular tachycardia. The primary clinical challenge lies in distinguishing between ventricular tachycardia (VT) and supraventricular tachycardia with aberrancy(SVT-A)\cite{kashou2021differentiating}. \autoref{fig:both} demonstrates the electrocardiographic overlap between VT and SVT-A, including shared features such as rapid heart rate and widened QRS complexes. This visual comparison reinforces the importance of applying validated algorithms to avoid misdiagnosis in acute settings. WCT should routinely be treated as VT(VT is significantly more common than SVT-A conduction, accounting for over $80\%$ of cases) unless the diagnosis of SVT-A or SVT with preexcitation is certain\cite{vereckei2014current}. Adenosine and other atrioventricular(AV) nodal–blocking agents are ineffective and potentially deleterious in patients with VT.\\
Numerous ECG criteria and interpretation algorithms have been developed to help clinicians distinguish VT from SVT-A\cite{swerdlow2001supraventricular},\cite{irusta2009algorithm},\cite{thomson1993automatic},\cite{antunes1994differential},\cite{griffith1994ventricular}. Till now the Brugada criteria remain the gold standard due to high sensitivity and guideline endorsement. Brugada et al. introduced a stepwise decision tree algorithm to differentiate between VT and SVT-A. The initial step evaluates all precordial leads (V1–V6) for the presence of RS complexes. VT is diagnosed, if no RS complexes are observed in any lead. If RS complexes are present, the longest RS interval in any precordial lead exceeding $100 ms$ is diagnostic of VT. After these steps, the presence of independent P waves, capture beats, or fusion beats also confirms VT. If the preceding steps are inconclusive, the diagnosis of VT is supported by the presence of a monophasic R, QR, or RS pattern with an R-wave duration greater than $40 ms$, or a notched S wave in lead V1. Additional supporting criteria include an R/S ratio less than $1$, QS or QR pattern in lead V6, an R-wave onset to S-wave nadir interval greater than $60 ms$ in leads V1 or V2, a notched S wave, or a delayed S-wave nadir (greater than $60 ms$), along with a QS or QR pattern in V6\cite{brugada1991new}. This algorithm reportedly achieved a sensitivity of $98.7\%$ and a specificity of $96.5\%$, outperforming traditional criteria. However, because of complex morphological criteria in its final steps, challenges in detecting AV dissociation at high heart rates and interpreter experience further affects accuracy. As a result, there is increasing interest in developing automated, computer-aided techniques for WCT differentiation to support clinicians in patient diagnosis and monitoring\cite{abdelmoneem2020arrhythmia}.\\
In recent years, deep learning (DL) algorithms—particularly convolutional neural networks (CNN)—have been widely adopted as automated, computer-aided tools for various biomedical applications, demonstrating promising performance\cite{tian2020deep}. DL networks are well-suited for ECG signal classification due to their capability to identify patterns and automatically learn relevant features directly from raw data\cite{alamatsaz2024lightweight}.\\
Although several approaches have been introduced for WCT differentiation using deep learning models, many rely on complex, deep architectures that may impede real-time processing. This highlights the need for lightweight DL models capable of delivering accurate WCT classification with minimal computational delay.\\
To counter the practical diagnostic limitations of manual and automated algorithms, as well as supplement their known strengths, we developed and validated a precise, lightweight automated diagnostic system designed to assist cardiologists by offering an interpretable deep learning approach that is time-efficient, cost-effective, reliable, and reduces the incidence of misdiagnosed VTs. The proposed model integrates CNN and recurrent neural networks (RNN), specifically long short-term memory (LSTM) units, to distinguish VT from SVT-A using $12$-lead ECG signals. In summary, the key contributions and innovations of this study are outlined as follows:

\begin{enumerate}  
\item Localized morphological features were extracted from individual ECG leads using dedicated CNN blocks. These modules employed convolutional layers to capture lead-specific patterns.
\item The derived feature vectors from each lead were concatenated into a unified representation. Subsequently, LSTM networks were applied to model spatial-temporal dependencies across leads, enabling the identification of inter-lead correlations critical for rhythm discrimination.
\item Leave-one-out cross-validation (LOOCV) was rigorously applied to partition training and testing datasets. This approach ensured strict separation of samples across folds, minimizing overfitting risks and preserving the integrity of model performance evaluation.
\item Shapley values are used to show the contribution or the importance of each ECG sample and lead to the prediction of the model.
\end{enumerate}  
Section \ref{wct} first describes the morphology of WCT. Section \ref{lr} reviews related work. The proposed system architecture is detailed in Section \ref{prp}. Section \ref{smrs} presents experimental results and compares them with state-of-the-art methods. Finally, Section \ref{con} discusses the current status of the proposed technique, along with its advantages and limitations.

\section{Morphology of Wide Complex Tachycardia}\label{wct}
In a healthy heart, electrical impulses typically originate from the sinoatrial (SA) node—commonly referred to as the heart’s natural pacemaker—and subsequently travel to the atrioventricular (AV) node. This electrical conduction triggers ventricular contraction, facilitating the forward movement of blood\cite{kashou2019physiology}. SVT and VT arise from aberrant cardiac signaling. In SVT, altered signaling triggers premature initiation of the heartbeat in the atria, whereas in VT, the premature onset occurs in the ventricles. These disruptions lead to an accelerated heart rate. Under normal resting conditions, the heart rate ranges from $60-100$ beats per minute (bpm). However, during episodes of SVT or VT, the heart rate exceeds $100 bpm$\cite{svt}. When a patient with a bundle branch block(BBB) experiences SVT, the result is a wide complex tachycardia(WCT) which becomes similar to VT as shown in \autoref{fig:both}. The similarities between VT and SVT-A are QRS duration $>120 ms$, regular rhythm, a heart rate $>100 bpm$, often ranging from $150–250 bpm$ and the clinical symptoms of both include palpitations, chest pain, dizziness, or shortness of breath\cite{vereckei2014current},\cite{garner2013wide},\cite{hampton2024ecg}. Accurate differentiation between VT and SVT-A is critical, as mistaking VT for SVT-A can result in delayed or inappropriate treatment of a life-threatening arrhythmia, while misdiagnosing SVT-A as VT may lead to unnecessary administration of antiarrhythmic drugs or cardioversion, posing avoidable risks to the patient. Differentiating between supraventricular SVT-A and VT is critical for guiding appropriate therapeutic interventions. Several factors aid in their distinction, including the patient’s clinical presentation, ECG characteristics, and response to specific pharmacological agents. Although diagnostic criteria exist to support a diagnosis of VT, no definitive criteria currently exist to reliably exclude VT\cite{alzand2011diagnostic}. Common diagnostic features suggestive of VT include generally regular R–R intervals (though not invariably so), an indeterminate atrial rate, and a ventricular rate typically ranging from $150-250 bpm$, The QRS complexes are characteristically wide (exceeding $120 ms$), lack preceding P waves, and the PR interval is not measured due to the ventricular origin of the rhythm. The QRS complexes often appear broad and abnormal in shape, with minimal distinction between the QRS complex and the T wave, making interpretation more challenging\cite{beed2014bennett}. For SVT-A, the rules are triphasic rsR’ in V1 for right BBB(RBBB), monophasic R in V1 for left BBB(LBBB), QRS duration $<140 ms$ (RBBB) or $<160 ms$ (LBBB), and absence of VT-specific signs like precordial concordance (all positive/negative QRS in V1–V6) or AV dissociation. Diagnostic algorithms (Brugada/Vereckei) prioritize the RS interval $<100 ms$ and lead-specific patterns (e.g., qRS in V6), while adenosine termination or clinical history (preexisting BBB, absence of structural heart disease) further support SVT-A\cite{brugada1991new}.

\section{Literature review}\label{lr}
To effectively distinguish VT from SVT-A, the development of optimized and precise automated detection algorithms is essential. This section discusses several of the most promising existing algorithms developed for the differentiation of these two cardiac signals. These algorithms fall into two categories, non-artificial intelligence(AI) and AI-based techniques. While there are existing algorithms that distinguish VT from SVT\cite{swerdlow2001supraventricular},\cite{sinha2004clinical},\cite{irusta2009algorithm},\cite{thomson1993automatic},\cite{ge2002cardiac},\cite{zheng2020optimal}, a critical diagnostic challenge lies in the subset of SVT with aberrancy (SVT-A), which closely mimics VT due to aberrant ventricular conduction. Aberrant conduction—caused by pre-existing BBBs, rate-dependent repolarization delays, or pre-excitation syndromes—generates wide QRS morphologies, RS interval prolongation, and pseudo ventricular patterns that overlap with VT features, rendering traditional criteria insufficient. This resemblance necessitates a targeted review of studies that explicitly analyze SVT-A vs. VT, rather than SVT broadly, to dissect how aberrancy blurs diagnostic boundaries. In $1991$, Brugada et al. presented a systematic approach to distinguish between VT, SVT with ectopic conduction, and SVT with synchronous conduction in a lateral pathway. The authors proposed a step-by-step algorithm based on specific electrocardiographic criteria. The duration of the QRS complex, the axis of the QRS complex in the frontal plane, the relationship between atrial and ventricular events, and the morphological characteristics of the QRS complex were evaluated as parameters. In this study, sensitivity and specificity were reported to be $98.7\%$ and $96.5\%$, respectively, for the diagnosis of VT\cite{antunes1994differential},\cite{brugada1991new}. While Brugada et al. proposed a highly sensitive ($98.7\%$) and specific ($96.5\%$) algorithm for differentiating VT from SVT-A, a subsequent study by Isenhour et al. challenged its real-world applicability by demonstrating significantly lower accuracy when the algorithm was used by board-certified emergency physicians and cardiologists. Isenhour et al. aimed to evaluate the Brugada algorithm's accuracy in a blinded fashion using a database of $157$ electrophysiologically proven WCTs. Their findings indicated that for cardiologist number 1, the sensitivity and specificity for VT diagnosis were $85\%$ and $60\%$, respectively. Cardiologist number 2 achieved a sensitivity of $91\%$ and a specificity of $55\%$. Emergency physician number 1 demonstrated a sensitivity of $83\%$  and a specificity of $43\%$, while emergency physician number 2 showed a sensitivity of $79\%$  and a specificity of $70\%$. These results markedly contrast with the original reported values and highlight that neither emergency physicians nor cardiologists could achieve the high levels of sensitivity and specificity reported by the Brugada study. This discrepancy suggests that while the Brugada algorithm may be theoretically robust, its practical application by a broader range of clinicians may be limited\cite{isenhour2000wide}. Vereckei et al. developed a new simplified algorithm to reduce reliance on complex morphological criteria. This diagnostic approach incorporates several key criteria: the presence of AV dissociation, the presence of an initial R wave in lead aVR, the correspondence of the WCT morphology to BBB or fascicular block, and the ventricular activation velocity ratio ($V_i / V_t$), where a ratio less than or equal to 1 suggests VT and a ratio greater than 1 suggests SVT-A. The study's findings revealed that the new algorithm demonstrated superior overall test accuracy of $90.3\%$ \cite{vereckei2007application}. In a completely different work, Badhwar et al. proposed using atrial overdrive pacing (AOD) to differentiate between VT and SVT-A. In a small cohort of eight patients with WCT and a 1:1 AV relationship, they paced the atrium and evaluated the change in QRS morphology during pacing and the first return electrogram after pacing cessation. The key results demonstrated that a change in QRS morphology during AOD indicated VT, while no change suggested SVT-A; furthermore, VT patients exhibited a ventricular response after pacing stopped, whereas SVT-A patients showed an atrial response, highlighting AOD as a potentially rapid and effective diagnostic tool despite the study's small sample size\cite{badhwar2009electrophysiological}. In another non-AI technique proposed in 2022, Reichlin et al. integrated clinical risk factors with specific ECG criteria to simplify the diagnostic process. The algorithm determined VT if two of the following three criteria were met: 1) high-risk clinical history, 2) time to peak in lead \textsc{ii} greater than $40 ms$, and 3) time to peak in lead aVR greater than $40 ms$. Their technique achieved $93\%$ sensitivity, $90\%$ specificity, and $93\%$ overall accuracy in validation tests.\cite{moccetti2022simplified}.
Traditional methods for diagnosing WCT rely on algorithms and physician expertise, often resulting in suboptimal accuracy, with reported rates up to $93\%$. Frontline physicians frequently struggle with these diagnoses due to variability in skill levels and limitations of existing algorithms.
To overcome existing limitations, Mazandarani et al. introduced dynamic time warping (DTW), a technique commonly used for measuring similarity between two temporal sequences that may vary in speed, to analyze WCT. On the testing subset of the dataset, the proposed method achieved a sensitivity of $97.46\%$, a specificity of $97.86\%$, and an overall accuracy of $97.66\%$ in discriminating between VT and SVT-A \cite{mazandarani2018wide}.\\ 
The integration of AI into cardiology is transforming diagnostic practices, particularly for interpreting complex cardiac rhythms. The AI techniques fall into two categories. Machine learning (ML) and DL techniques. In ML techniques the features are extracted manually but DL methods are end-to-end techniques. In a study conducted by Zhen-Zhen Li and colleagues, a gradient boosting machine (GBM) model was used and demonstrated superior performance over traditional methods, achieving an accuracy of $94.44\%$ in distinguishing VT from SVT-A in WCT cases. This suggests that ML models could significantly aid in clinical decision-making by overcoming the limitations of conventional diagnostic algorithms\cite{li2024machine}. Recent advancements in the use of CNNs, have demonstrated improved performance by automatically extracting and analyzing features from ECG data. The study presented by Fayazifar et al. focuses on the development and rigorous evaluation of a novel CNN model. A key innovation in their approach was the utilization of a neural architecture search(NAS) method, specifically differentiable architecture search(DARTS), to autonomously discover an optimized neural network structure. The resulting architecture comprised a stem convolution layer followed by five specialized cells, incorporating separable-convolution and dilated-separable-convolution layers, arranged in an N-R-N-R-N sequence (Normal-Reduction cell pattern). The model achieved an accuracy of $91.1\%$, a sensitivity of $90\%$, a specificity of $90\%$, and an F1-score of $88.67\%$. A notable advantage of this CNN approach is its ability to automatically extract salient features directly from raw ECG signals, bypassing the need for manual noise filtering or feature engineering, a common bottleneck in traditional ML methods\cite{fayyazifar2023novel}. In 2024, Chow et al. built on this foundation by developing and testing a novel CNN architecture specifically designed for WCT classification using 12-lead ECGs, aiming to address diagnostic challenges with greater speed and accuracy than human cardiologists. Their proposed model used CNN with DARTS optimization and had superior sensitivity and accuracy of $91.9\%$ and $93.0\%$, respectively than non-EP cardiologists and similar performance compared with EP cardiologists \cite{chow2024interpreting}. 
In a 2024 study led by Adam M. May, the authors demonstrated that automated algorithms generally outperformed traditional methods in terms of accuracy and specificity, highlighting their potential to enhance clinical decision-making in WCT diagnosis\cite{lococo2024direct}.

\section{Wide Complex Tachycardia classification methodology}\label{prp}

In the WCT discrimination approach outlined in this study, an initial pre-processing phase is conducted to prepare the data for input into the deep neural network (DNN) algorithm, followed by the execution of training and testing procedures using the DNN algorithm. The pre-processing phase encompasses the following steps: noise removal, normalization, and data segmentation.
\subsection{ECG Dataset}
This study compiled a database comprising ECG recordings from $35$ patients referred to the electrophysiology (EP) laboratory at Rajaei Cardiovascular, Medical, and Research Center from $2014$ to $2016$. The database included one-minute 12-lead long-term ECG excerpts for each participant. The WCT diagnoses were confirmed by two certified electrophysiologists using standardized electrophysiological protocols, with annotations applied to each recording. Data acquisition was performed at a sampling rate of  $\SI{1000}{\hertz}$ per channel, with 16-bit resolution across a $10 mV$ dynamic range. Electrophysiological analysis revealed WCT diagnoses of SVT-A in $24$ patients and VT in $11$ cases.\\
In this study, leave-one-out cross-validation (LOOCV) was implemented to assess the robustness of the proposed algorithm. LOOCV, a rigorous resampling technique, operates as a specialized case of k-fold cross-validation where k corresponds to the total number of patients in the dataset \cite{berrar2019cross}. For each iteration, one observation from the dataset was reserved for validation, while the remaining observations were utilized for training. This process was repeated exhaustively until every sample in the dataset had served as the validation instance once. Despite its computational intensity, LOOCV was selected for its capacity to minimize bias in performance estimation—a critical consideration given the limited sample size of the dataset. 

\subsection{Preprocessing}
 \subsubsection{Noise filtering and Normalization}
ECG signals are often contaminated by various forms of noise, including low-frequency and high-frequency disturbances such as baseline wander (BW), power line interference, electromyographic (EMG) noise, and electrode motion artifacts. To mitigate these interferences, different filtering techniques can be employed. BW, a low-frequency artifact in ECG recordings, is primarily induced by respiration and subject movement. We employ a bandpass filter with a cutoff frequency of $\SI{140}{\hertz}$ to mitigate the impact of artifacts in the recordings. Before subsequent analysis, it is essential to account for variations in the amplitude of the ECG signal. To address these variations, which arise from differences in instrumentation and individual subjects across recordings, we apply a normalization technique. Specifically, to eliminate the offset effect, the mean value of each record is subtracted from each sample, followed by division by the record’s standard deviation. This normalization process ensures that each record exhibits a mean of zero and a standard deviation of one.

\subsubsection{Segmentation} 
Segmentation of ECG recordings standardizes the data length before input into the model. With the sampling rate of $\SI{1000}{\hertz}$ and an average cardiac cycle of $300–400 ms$ for VT and $250–400 ms$ for SVT-A, segments with $500$ samples ($500 ms$) seem appropriate, since at least one WCT beat appears within this length. To segmentize the ECG records, we utilized the pre-trained model developed by Cai et.al to detect R peaks\cite{cai2020qrs}. This pre-trained CNN model is mainly composed of convolutional blocks and squeeze-and-excitation networks(SENet). Each window spans $250$ samples around the R-peak, ensuring the fiducial point is aligned at the midpoint. The sliding window traverses through the records beat by beat and produces the discrete sections. After this step, all ECG segments from all patients are combined. As shown in \autoref{tab1} number of segments related to VT and SVT-A classes was comparable resulting in a balanced dataset.


\begin{table}[h]
\centering
\caption{Number of each rhythm after segmentation}
\setlength{\tabcolsep}{3pt}
\begin{tabular}{|m{60pt}|c|}
\hline
WCT type&	Total number of segments\\
\hline

SVT-A&3459\\
\hline
VT&2798\\
\hline

\end{tabular}
\label{tab1}
\end{table}

\subsection{Proposed model architecture}
This study proposes a highly accurate deep learning approach that integrates CNN and LSTM architectures for each lead, enabling end-to-end differentiation of VT from SVT-A using 12-lead ECG signals.\\
\begin{figure}[htbp]
\centerline{\includegraphics[width=0.75\columnwidth]{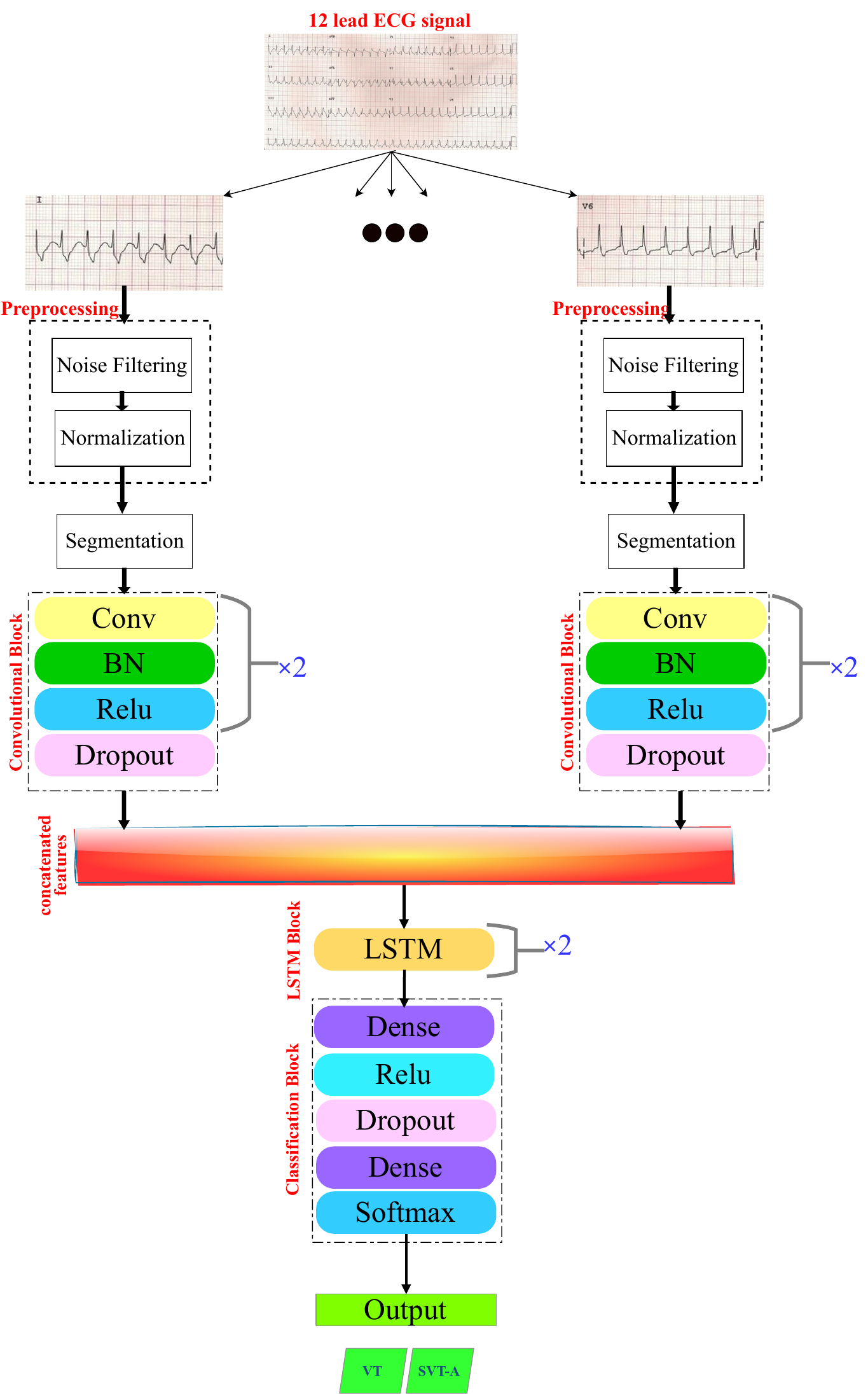}}
\caption{Architecture of the proposed deep learning model}
\label{fig1}
\end{figure}
The architecture of the proposed model as illustrated in \autoref{fig1} consists of 2 blocks of lead-specific CNN feature extraction for each lead and 2 LSTM layers deployed on concatenated features and 2 fully connected layers to classify. The end-to-end model is trained and evaluated using a leave-one-out validation technique. The architecture of the proposed model as shown is as follows: For each lead 2 tandem convolutional blocks, each one contains a 1-dimensional convolutional layer for extracting local features and a batch normalization layer which normalizes the inputs of the next layer to accelerate the convergence and an activation function, called rectified linear unit (ReLU) for achieving non-linear capabilities. The output of the last ReLU layers is then given to the dropout layer that deactivates $20\%$ of neurons in that layer for each batch to prevent overfitting and Improve robustness.  The outputs of the last convolutional block of each lead are then concatenated to produce a sequential input to be fed to an LSTM layer to extract spatial dependencies between them. The LSTM architecture incorporates memory blocks and recursive feedback connections, enabling it to effectively manage long-term dependencies and mitigate issues related to exploding gradients. In this study, two stacked LSTM layers were employed to process sequential features. Fully connected layers, typically positioned at the network’s final stages, were utilized to adjust the dimensionality of the input vector, transforming low-dimensional features into linear representations. The softmax activation function was applied in the final layer to convert the output vector into class probabilities for binary classification. Given the categorical nature of the target variable, one-hot encoding was implemented during the classification process. \autoref{tab2} gives a detailed description of the proposed model.\\ In DNNs, two categories of hyperparameters require optimization before the training phase: those associated with the network architecture, including dropout rate, weight initialization, and activation function, and those related to the training algorithm, such as learning rate, momentum, batch size, and the number of epochs.\cite{montavon2012neural}. To identify the optimal hyperparameter configuration for maximizing accuracy, we employed a grid search technique. In our study, the optimal number of hidden layers and neurons was determined manually. This approach is built upon the foundational work established in prior research\cite{nasimi2022ldiaed}. We iteratively reduced the number of layers until a decline in accuracy was observed. Employing a grid search technique to optimize batch size and the number of epochs yielded values of $32$ and $100$, respectively. This method also identified the "Adam" optimization algorithm as the most effective for iteratively updating network weights during training. The learning rate, a key hyperparameter, was set to $0.001$ and decreased by a factor of $10$ when the validation loss ceased to improve for two consecutive epochs. To determine the optimal network weight initialization, we evaluated all available techniques using grid search, ultimately selecting "he\_normal." For tuning the dropout hyperparameter, a range of $0.0 - 0.9$ was explored via grid search, resulting in a dropout rate of $0.2$ to mitigate overfitting.
\begin{table}[!h]
\centering
\caption{ Detailed description of the proposed deep learning model }
\setlength{\tabcolsep}{3pt}
\begin{tabular}{|c|c|c|c|c|}

\hline
Layers	&Filter size&	Kernel size&	Strides& Output dimension\\
\hline
Conv1D&	32&	16&1&(500,32)\\
\hline
BatchNorm &\multicolumn{3}{c|}{}&(500,32)\\	
\hline
Relu &\multicolumn{3}{c|}{}&(500,32)\\	
\hline
Conv1D	&32&	16&4&(125,32)\\
\hline
BatchNorm &\multicolumn{3}{c|}{}&(125,32)\\	
\hline
Relu &\multicolumn{3}{c|}{}&(125,32)\\	
\hline
Dropout&	 \multicolumn{3}{c|}{Dropout rate = 0.2}&(125,32)\\
\hline
Concatenated &\multicolumn{3}{c|}{}&(125,384)\\	
\hline
LSTM	&\multicolumn{3}{c|}{units=128}&(125,128)\\
\hline
LSTM	&\multicolumn{3}{c|}{units=64}&(125,64)\\
\hline
Dense	&\multicolumn{3}{c|}{}&(128)\\
\hline
Relu &\multicolumn{3}{c|}{}&(128)\\	
\hline
Dropout&\multicolumn{3}{c|}{	Dropout rate = 0.2}&(128)\\
\hline
Dense	&\multicolumn{3}{c|}{}&(2)\\
\hline
softmax &\multicolumn{3}{c|}{output=2}&(2)\\	
\hline
\end{tabular}
\label{tab2}
\end{table}

\section{Experiments}\label{smrs}
The research conducted aimed to differentiate WCT using a DL model.  We employed LOOCV to rigorously evaluate model performance while preventing data leakage. To optimize performance and prevent overfitting, we trained the model for a maximum of $100$ epochs but implemented early stopping based on validation loss. Training typically halted between epochs $20-30$ when no further improvement in validation loss was observed (patience = 2 epochs). Categorical cross-entropy was employed as the loss function, with the "Adam" optimizer selected due to its ability to accelerate algorithm convergence, ease of implementation, computational efficiency, and the effectiveness of its default hyperparameters with minimal tuning. Weights and biases were iteratively updated using a conjugate gradient back-propagation algorithm, which leverages the loss function’s output to optimize these parameters until the desired values are achieved. Upon completion of the classification process, Shapley value analysis was conducted for the target classes, utilizing the methods outlined in\cite{Lundberg2021}. This research is carried out in Python programming language in JupyterLab from the Anaconda distribution of Python 3.12.0 on a system with an Intel Core i7 7th Gen processor with NVIDIA GeForce GTX 1070 Ti.

\subsection{Experimental results and discussion }
\begin{figure}[h]
\centerline{\includegraphics[width=0.5\columnwidth]{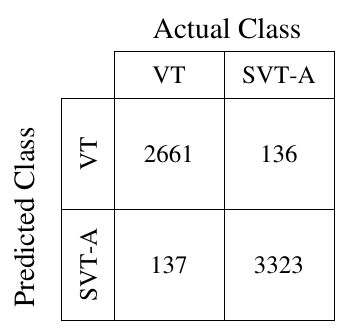}}
\caption{Aggregated confusion matrix }
\label{fig2}
\end{figure}

To evaluate our proposed model, we need a foundational tool for assessing. Confusion matrix is indeed this tool, but before using it, we need to assign positive and negative classes to these two WCTs. In the context of binary classification for differentiating VT from SVT-A, the assignment of positive and negative classes depends on the clinical goal and what we want to detect. Because VT is a life-threatening arrhythmia and requires immediate intervention, and missing VT could lead to death, the goal is to maximize the detection of VT cases, so this class should be positive. On the other side, SVT-A is less immediately dangerous and often is treated with medications or less urgent interventions, and misclassifying SVT-A as VT may lead to unnecessary treatment(e.g. inappropriate shocks), but this is generally considered less harmful than missing VT. \\A confusion matrix provides a detailed breakdown of model predictions: In this context, true positives(TP) means the number of VTs correctly classified as VT. True negatives(TN) means the number of SVT-As correctly predicted as SVT-A. False positives(FP) mean the number of SVT-As misclassified as VT and false negatives(FN) mean the number of VTs misclassified as SVT-A. As mentioned earlier LOOCV was employed to evaluate all experimental scenarios. The model was configured to output instances of FP, FN, TP, and TN. These classification outcomes were aggregated across all test scenarios to construct a comprehensive confusion matrix ( \autoref{fig2}).  Numbers on the main diameter of this matrix indicate correct classification, while off-diagonal cells show misclassifications.

\begin{figure}[h]
\centerline{\includegraphics[width=\columnwidth]{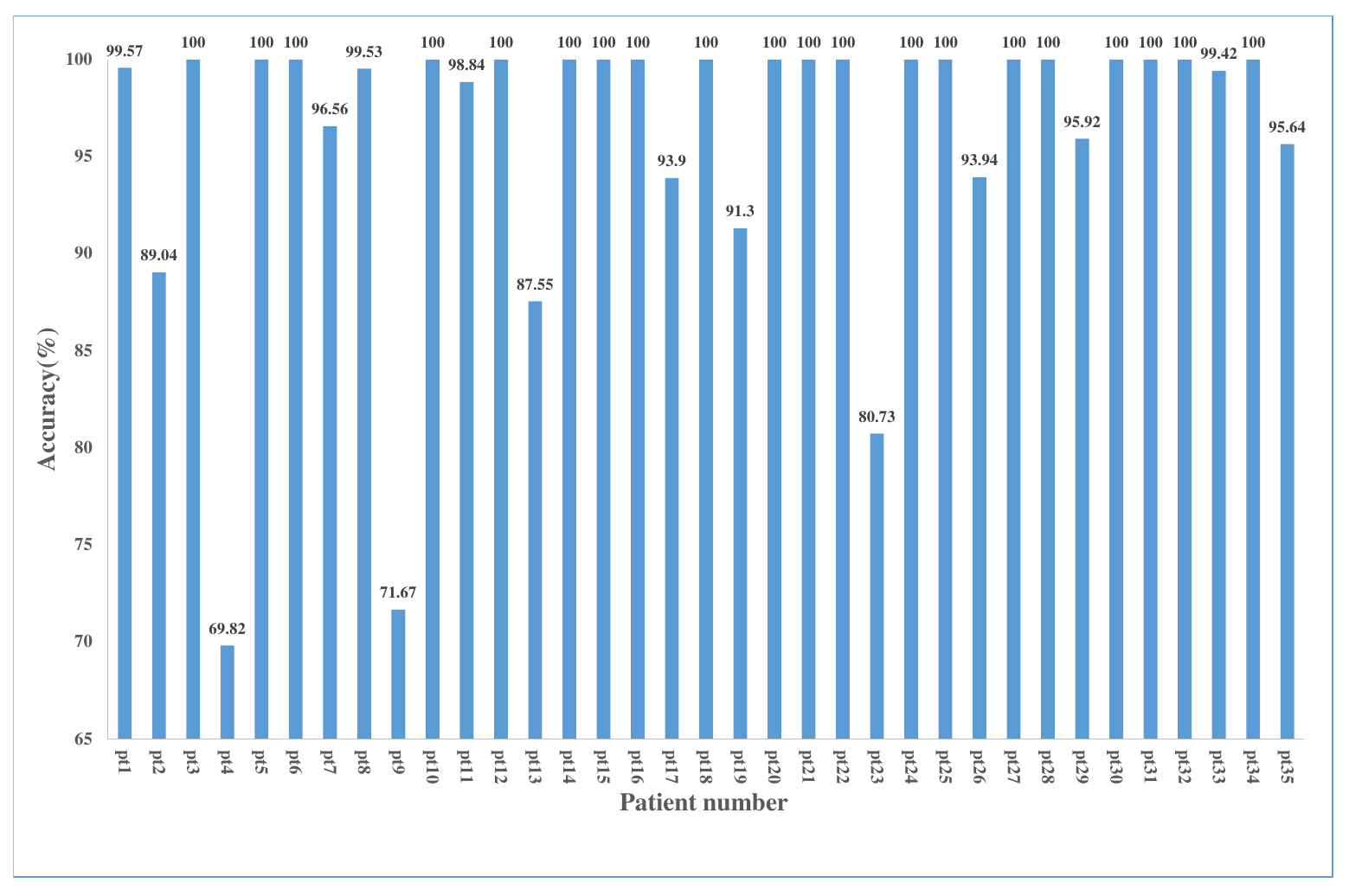}}
\caption{Accuracy of the proposed technique for each patient}
\label{acc}
\end{figure}

\begin{figure}[h]
\centerline{\includegraphics[width=\columnwidth]{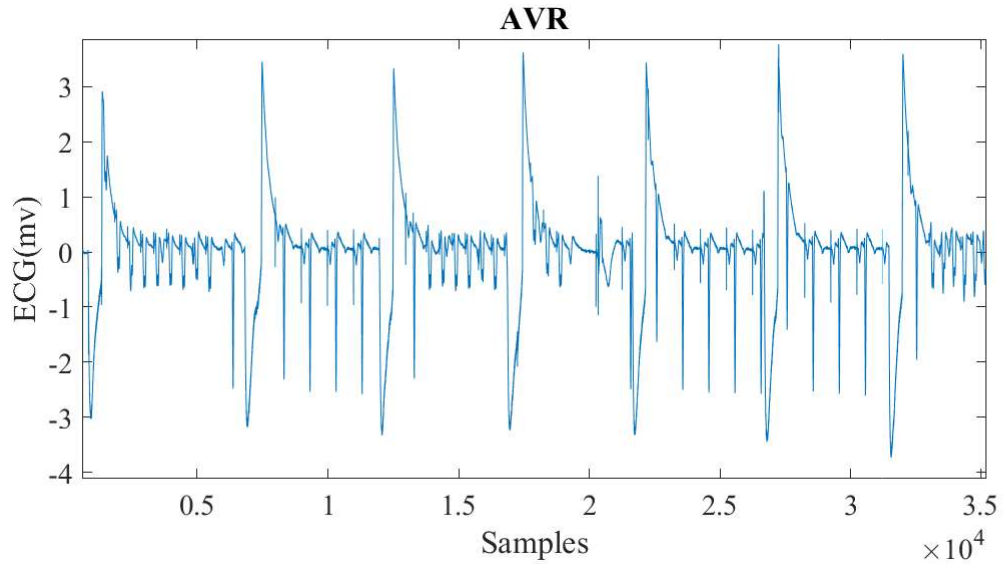}}
\caption{Noisy ECG signal(aVR) related to patient number 4.}
\label{vt10}
\end{figure}

The confusion matrix is essential for diagnosing model behavior, but one has to use derived metrics like sensitivity, specificity, accuracy and, F1-score which are actionable summaries tailored to specific goals. Sensitivity is the ability of the proposed model to detect VT in patients who truly have VT. In $95.10\%$ of cases our model can truly detect VT; missing $4.90\%$ of true VT cases. This parameter is critical in life-threatening scenarios like VT diagnosis. Specificity is the ability to rule out VT in patients with SVT-A, which is $96.06\%$ in our model, meaning that the model incorrectly flags $3.94\%$ of SVT-A cases as VT. F1-score($95.12\%$) is a harmonic mean of precision and sensitivity. \autoref{acc} shows the accuracy metric across all tested scenarios.\\
Accuracy shows the overall correctness of the model across both classes. In other words, indicates how accurately the model has performed. The classification accuracy achieved by the proposed model was $95.63\%$ with a $95\%$ confidence interval of [$93.07\%$, $98.19\%$] meaning that the overall correctness or reliability of the proposed model is $95.63\%$ and With $95\% $confidence that the accuracy of distinguishing the VT from the SVT-A for new data by the presented model will be in the range of $93.07\%$ to $98.19\%$(A narrower confidence interval indicates a more precise estimate, meaning there is less uncertainty about the true population accuracy value).\\
As one can see in \autoref{acc} there are three patients whose diagnostic accuracy is below $80\%$ and have the highest misclassification rate. For example from $338$ VT test segments related to patient number $4$, $102$ segments were misclassified as SVT-A. By diving deeper into the main record of the mentioned patient as shown in \autoref{vt10} we recognize that the ECG signal under consideration is a noisy recording, which poses challenges in definitively determining the accuracy of annotations made by cardiologists and/or the algorithm.\\

The parallel lightweight model implemented in this study revealed a high-performance accuracy for the differentiation of VT from SVT-A. Unlike conventional deep learning frameworks that employ extensive convolutional hierarchies, our model uses only two convolutional blocks per lead to extract lead-specific morphological features, and at the end uses two fully connected layers for global classification. This minimalist architecture prioritizes parameter efficiency and interpretability, ensuring compatibility with small datasets while retaining diagnostic fidelity. By avoiding excessive abstraction, the model preserves electrophysiological patterns critical to differentiating SVT-A from VT, as validated by accuracy performance comparable to state-of-the-art, resource-intensive models. The performance comparison with other research studies about the same problem is given in \autoref{tab3}.
In comparison to the established Brugada algorithm—which has historically demonstrated high performance in WCT diagnosis—our proposed AI-based method exhibits comparatively lower sensitivity. However, the real-world applicability of the Brugada algorithm has been called into question. As reported by Isenhour et al.~\cite{isenhour2000wide}, the diagnostic sensitivity significantly decreased when the algorithm was applied by board-certified emergency physicians and cardiologists. Specifically, for one group of cardiologists, the sensitivity and specificity were reported as $85\%$ and $60\%$, respectively. Moreover, the Brugada algorithm involves a complex set of rule-based criteria, which may limit its practicality and consistency in time-sensitive clinical settings. As shown in the table below, our proposed method outperforms the approaches presented in \cite{vereckei2007application}, \cite{li2024machine}, \cite{fayyazifar2023novel}, and \cite{chow2024interpreting}. When compared to the non-AI method proposed in \cite{mazandarani2018wide}, our approach—despite being evaluated on the same dataset—yields comparatively lower performance. However, the proposed AI-based framework offers advantages in terms of scalability, interpretability, and automation, making it more suitable for future deployment and continuous improvement as more data becomes available.

\begin{table}[htb]
\caption{ Summary of VT vs. SVT-A discrimination techniques}
\centering

\setlength{\tabcolsep}{2.5pt}
\begin{tabular}{|m{60pt}|m{70pt}|c|c|c|c|}

\hline
Ref(work)&Classification technique&	Accuracy&Sensitivity&Specificity&F1-score\\
\hline
Brugada et al.\cite{brugada1991new}&Non-AI&-&98.7\% & 96.5\%&-\\ \hline 
Vereckei et al.\cite{vereckei2007application}&Non-AI&90.03\% &	-&-&-\\ \hline 
Niknejad Mazandarani et al.\cite{mazandarani2018wide}&Non-AI&	97.66\% 	&97.46\%&97.86\%&	-\\ \hline  
Li et al.\cite{li2024machine}&AI(ML) &	94.44\%&	-&-&-\\ \hline 
Fayyazifar et al.\cite{fayyazifar2023novel}&AI(DL)&91.1\%&	90\%& 90\%&88.67\%\\ \hline  
chow et al.\cite{chow2024interpreting}&AI(DL)&93\%&	91.9\%& -&-\\ \hline 
\textbf{Proposed}&\textbf{AI(DL)}&\textbf{95.63\%}&\textbf{95.10\%}&\textbf{96.06\%}&\textbf{95.12\%}\\

\hline
\end{tabular}
\label{tab3}

\end{table}

\subsection{Explainability of the DL model}
ML and DL models often function as opaque "black boxes", creating interpretability challenges that undermine clinical confidence in their predictions. To establish trustworthiness, we implemented SHAP (SHapley Additive exPlanations) values for both local and global explanations. Local SHAP quantifies feature(sample)-level contributions to individual predictions, while global SHAP aggregates these local explanations to identify the most influential ECG leads across the entire dataset. So we used the functions proposed in \cite{Lundberg2021} to show the contribution or the importance of each feature (sample) and each lead on the prediction of the model. 
To identify the most influential features contributing to the model's predictions, we applied a statistical thresholding approach. Specifically, we computed the mean ($\mu$) and standard deviation ($\sigma$) of the SHAP values across each lead and defined a threshold as:
\[
\text{Threshold} = \mu + \sigma
\]

Features with SHAP values greater than or equal to the threshold were considered significantly impactful:

\[
\text{SHAP}_i \geq \mu + \sigma
\]

\begin{figure}[htbp]
\centering
  \begin{subfigure}[h]{0.9\textwidth} 
\centering
    \includegraphics[width=0.9\textwidth]{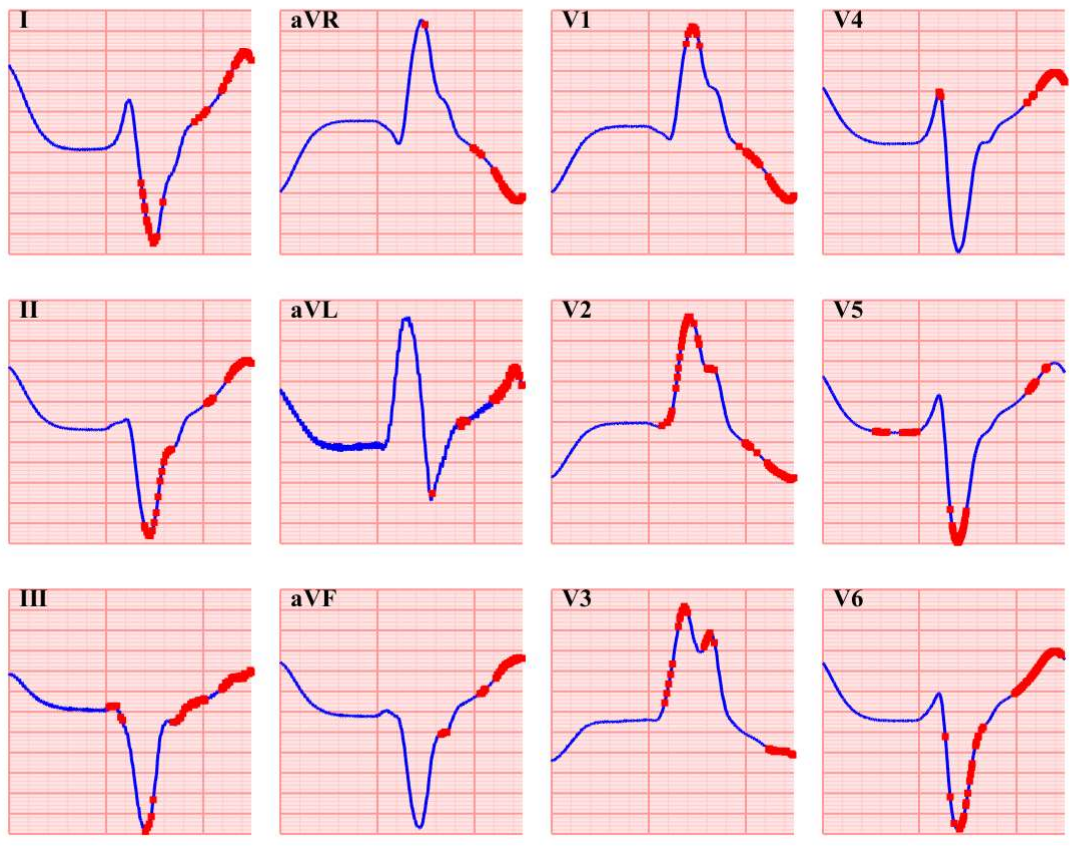}

    \label{fig:top}
  \end{subfigure}
  
  \vspace{0.5cm} 
  
  \begin{subfigure}[h]{0.9\textwidth} 
\centering
    \includegraphics[width=0.9\textwidth]{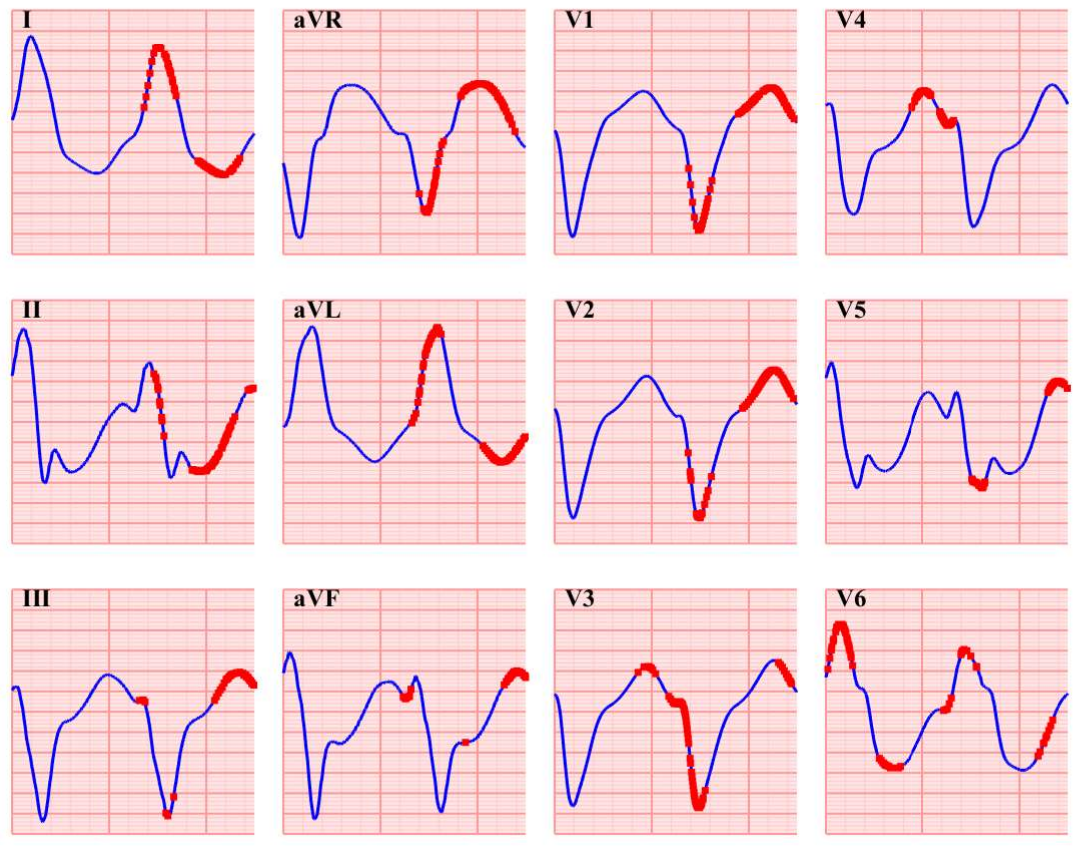}
    \label{fig:bottom}
  \end{subfigure}
  
  \caption{Top)Shapley Values Analysis for VT Class. Bottom)Shapley Values Analysis for SVT-A Class.} 
  \label{shap}
\end{figure}

This criterion prioritizes features that exhibit higher-than-average importance for model decision-making, thereby improving interpretability by focusing on the most relevant explanatory variables. Samples with higher Shapley values indicate a greater influence on the model's predictions, whereas those with lower values suggest minimal impact. To evaluate the Shapley values, we utilized the subset of the dataset that the model had not previously seen. Globally, SHAP facilitated a lead-specific relevance analysis for each WCT, directing clinical experts to concentrate on the most impactful leads identified. The findings underscored the V6 lead as the most critical in distinguishing WCT. This insight is particularly valuable for researchers studying specific arrhythmia predictions, as it highlights the importance of prioritizing the V6 lead in ECG monitoring. We have presented a visual depiction of the Shapley values assigned to various samples and leads of ECG signals. \autoref{shap} corresponds to a specific class and in each figure red dots correspond to the top Shapley values(influential samples in determining the prediction). 

\section{Conclusion and Future work}\label{con}
Accurate differentiation between VT and SVT-A is critical in emergency cardiac care, as the therapeutic strategies differ significantly. Misclassification may lead to inappropriate treatment, posing a risk to patient safety. In this study, we developed a lightweight deep learning-based classification model designed specifically to distinguish VT from SVT-A using 12-lead ECG recordings acquired from the Rajaei Cardiovascular, Medical, and Research Center. Our approach achieved an overall classification accuracy of $95.63\%$, demonstrating strong potential for clinical application, particularly in real-time decision support scenarios. The model architecture integrates CNNs to extract morphological and spatial features from ECG waveforms and LSTM layers to capture temporal dependencies across beats. This combination enhances the model's ability to capture subtle, clinically relevant differences between VT and SVT-A, which are often difficult to discern through manual analysis alone.
To ensure the interpretability of the model, SHAP values were employed to provide both global and local explanations for classification decisions. This transparency is essential in medical applications, where clinicians must understand and trust the output of automated diagnostic tools. The integration of SHAP contributes to a more explainable AI system, potentially improving clinician confidence and adoption in practice.\\
\textbf{Future work} will focus on several key directions. First, expanding the dataset with a more diverse patient ECGs across multiple institutions would improve generalizability and robustness. Second, integrating additional physiological signals such as heart rate variability or patient metadata (e.g., age, history of arrhythmia) could further enhance performance. Third, combining rule-based algorithms or hybrid systems with the current deep learning model may offer improved reliability, particularly in borderline or noisy cases. Finally, validating the model prospectively in clinical settings and integrating it into decision-support systems will be crucial steps toward real-world deployment.
Ultimately, while AI-powered tools offer promising assistance in ECG interpretation, final clinical decisions should remain under the judgment of qualified healthcare professionals, who can interpret algorithmic output within the broader clinical context.

\bibliographystyle{IEEEtran}
\bibliography{refrences}

\end{document}